# Super Space-time, a Model for Gravity and Dark Matter

Nasser Boroojerdian

Department of Mathematics and Computer Science,
Amirkabir University of Technology, Tehran, Iran

Email:broojerd@aut.ac.ir

**Abstract**

In this paper, we use the notion of super tangent bundle for a space-time manifold, and use super metrics on it to model gravity and some types of matter that is the source of the gravity and can be interpreted as dark matter. In this view, gravity is an effect of the symmetric part of the super metric and dark matter is an effect of the symplectic part of the super metric. So, dark matter can be considered as a result of the odd part of the super space-time. We use variation of the Hilbert action to derive field equations for interaction of gravity and matter and investigate some of the results. We also construct some solution for these equations.

1.Introduction

Many essential features of physical phenomena are described by superstructures [5][6]. It seems the basic mathematical structures for modeling matter and space relies on super symmetry. Our approach to work with super structures on manifolds is to construct a super vector bundle on manifolds whose even subbundle is its tangent bundle. And define a super Lie algebra structure on the sections of this super vector bundle whose even Lie algebra is the ordinary vector fields with ordinary Lie bracket [3].

In General Relativity, gravity is described by a manifold with a semi-Riemannian metric on it that is called space-time manifold. We can extend this structure to a super structure and construct a super space-time that is equipped with a super metric. Even part of this extension contains ordinary space-time and gravity is part of this structure. But also, odd part of this structure contains a new ingredient that resembles some sort of the matter that has only gravitational effect. So, it is a good candidate for dark matter. In this view, gravity is an effect of the even part of the super space-time and dark matter is an effect of the odd part of the super space-time.

In section 2 we remind some basic notions about super vector spaces and tensors and super inner products. In section 3, we introduce the notion of super tangent bundles and super Lie algebra structure of its sections and super differential forms, as developed in [3]. In section 4 we introduce the notion of super connections and super metrics on super tangent bundles and associated concepts such as curvatures. In section 5 we find explicitly curvatures of a special super tangent bundle with respect to the curvatures of even part of the super metric and data of odd part of the super metric. In section 6, we apply the results of pervious sections, for constructing a super space-time. In this section, by using variation of scalar curvature of the super metric [1][2], we find some equations that relate Ricci curvature to the some tensor that comes from odd part of the super metric. These equations are field equations that describe gravity and its sources. In section 7 we find some solutions for these equations.

## 2. Preliminaries

We have prepared the basic notions we need in this paper in [3]. We use conventions and concepts about super spaces and tensors on super spaces as described in [3]. In a super space $V = V_0 \oplus V_1$ the parity of a homogeneous element $x$ is denoted by $|x|$ as an element of $\mathbb{Z}_2$. For simplicity, we denote $(-1)^{|x|}$ by $(-1)^x$. Elements of $\otimes^k V^*$ are covariant tensors on $V$ of order $k$ that can be represented as $k$-linear maps $\widetilde{\omega}: V \times \cdots \times V \to \mathbb{R}$. These tensors are called even or odd, depending on preserving parities of vectors or reversing parities.

A tensor $\widetilde{\omega}$ of order $k$, is called super alternating, if for homogenous vectors $v_1, \cdots, v_k$, and indices $i = 1, \cdots, k-1$

$$\widetilde{\omega}(v_1, \cdots, v_i, v_{i+1}, \cdots, v_k) = -(-1)^{v_i v_{i+1}} \widetilde{\omega}(v_1, \cdots, v_{i+1}, v_i, \cdots, v_k) \quad (1)$$

And $\widetilde{\omega}$ is called super symmetric, if for homogenous vectors $v_1, \cdots, v_k$, and indices $i = 1, \cdots, k-1$

$$\widetilde{\omega}(v_1,\cdots,v_i,v_{i+1},\cdots,v_k) = (-1)^{v_i v_{i+1}} \widetilde{\omega}(v_1,\cdots,v_{i+1},v_i,\cdots,v_k) \qquad (2)$$

**super inner product**: A super inner product on a super space $V$, is a nonsingular bilinear map on $V$, such that:

$$\begin{aligned} <u,v> &= (-1)^{uv} <v,u> \\ u \in V_0, v \in V_1 &\Rightarrow <u,v> = 0 \end{aligned} \qquad (3)$$

This product is an (ordinary) inner product on $V_0$ and is a nonsingular symplectic product on $V_1$.

For every homogeneous bases $\{e_i\}$ of $V$ there exists a unique base $\{e^i\}$ such that $<e^j, e_i> = \delta_i^j$. The base $\{e^i\}$ is homogenous too, and $|e_i| = |e^i|$. The base $\{e^i\}$ is called the reciprocal base of $\{e_i\}$. The reciprocal base of $\{e^i\}$ is $\{(-1)^{e_i} e_i\}$. The dual base of $\{e_i\}$ is also denoted by $\{e^i\}$.

**Super contraction of tensors:** Suppose $T: V \times V \to \mathbb{R}$ is a bilinear map that preserves parity. The super contraction of $T$ is a scalar, corresponds to the super trace of linear maps. Consider a homogenous base $\{e_i\}$ and its reciprocal $\{e^i\}$. The super contraction of $T$ is defined as follows.

$$\text{Super contraction of } T = \text{sc}(T) = \sum_i T(e_i, e^i) \qquad (4)$$

### 3. Super tangent bundles

In [3], we developed the notion of super tangent bundle. Let $W$ be a vector space and $M$ be a manifold. Define $\widehat{T}M = TM \oplus (M \times W)$. This is a super vector bundle that its even subbundle is $TM$ and its odd subbundle is the trivial vector bundle $M \times W$. Even sections of this bundle are ordinary vector fields on $M$, and odd sections, are $W$-valued smooth functions on $M$. All sections of $\widehat{T}M$ are called super vector fields on $M$ and is denoted by $\widehat{\mathfrak{X}}M$. This is a super space that its even part is $\mathfrak{X}M$ and its odd parts is $C^\infty(M,W)$, so $\widehat{\mathfrak{X}}M = \mathfrak{X}M \oplus C^\infty(M,W)$. We use symbols $U, V, \cdots$ for ordinary vector fields, and $\alpha, \beta, \cdots$ for odd vector fields, and $\boldsymbol{U}, \boldsymbol{V}, \ldots$ for arbitrary vector fields on $M$. An arbitrary vector filed has the form $U + \alpha$ and Lie bracket of two super vector field $U + \alpha$ and $+\beta$, is defined as follows.

$$[U + \alpha, V + \beta] = [U,V] + U(\beta) - V(\alpha) \tag{5}$$

This definition make $\widehat{\mathfrak{X}}M$ into a super lie algebra that its even subalgebra is $\mathfrak{X}M$.

For every $f \in C^\infty(M)$ and odd field $\alpha$ it is defined $\alpha(f) = 0$. For $U, V \in \widehat{\mathfrak{X}}M$ and $f \in C^\infty(M)$ the following relation holds.

$$[U, fV] = U(f)V + f[U,V] \tag{6}$$

**Super differential forms**: For a super bundle $E$, a covariant $E$-valued (super) tensor field on $M$ can be defined as an operator that is $C^\infty(M)$-multilinear.

$$\omega: \widehat{\mathfrak{X}}M \times \cdots \times \widehat{\mathfrak{X}}M \longrightarrow \Gamma E \tag{7}$$

If $\omega$ preserve parities of vectors, it is called even, and if it reverse parities of vectors, it is called odd. For special case $E = M \times \mathbb{R}$, $E$-valued tensors are called numerical valued tensors.

The tensor field $\omega$ of order $k$, is called super alternating, if

$$\omega(U_1, \cdots, U_i, U_{i+1}, \cdots, U_k) = -(-1)^{U_i U_{i+1}} \omega(U_1, \cdots, U_{i+1}, U_i, \cdots, U_k) \tag{8}$$

These tensors are called (super) differential forms. The set of all $E$-valued (super) $k$-differential forms is denoted by $\hat{A}^k(M, E)$. In the case of $E = M \times \mathbb{R}$, the set of all numerical valued (super) $k$-differential forms is denoted by $\hat{A}^k(M)$.

Super symmetric tensor fields are defined similarly, as follows.

$$\omega(U_1, \cdots, U_i, U_{i+1}, \cdots, U_k) = (-1)^{U_i U_{i+1}} \omega(U_1, \cdots, U_{i+1}, U_i, \cdots, U_k) \tag{9}$$

4. **Super connections and super metrics**

A super connection (or super covariant derivation) on $M$ is a bilinear map (with respect to scalars)

$$\nabla: \widehat{\mathfrak{X}}M \times \widehat{\mathfrak{X}}M \longrightarrow \widehat{\mathfrak{X}}M, \quad (U, V) \mapsto \nabla_U V \tag{10}$$

such that the following relations hold.

1) $|\nabla_U V| = |U| + |V|$  (11)
2) $\nabla_U(fV) = U(f)V + f\nabla_U V$  (12)
3) $\nabla_{fU} V = f\nabla_U V$  (13)

The restriction of $\nabla$ to $\mathfrak{X}M \times \mathfrak{X}M$ is an ordinary connection on $M$, and its restriction to $\mathfrak{X}M \times C^\infty(M,W)$ is a connection on the trivial bundle $M \times W$. So, for some $\omega_0 \in A^1(M, L(W))$, we have:

$$\nabla_U \alpha = U(\alpha) + \omega_0(U, \alpha) \quad (14)$$

Moreover, for every odd field $\alpha$ the operator $\omega_\alpha: U \mapsto \nabla_\alpha U$ is a $W$-valued one form on $M$. In other words, $\omega_\alpha \in A^1(M, W)$, furthermore, $\alpha \mapsto \omega_\alpha$ is linear. For any two odd fields $\alpha, \beta$, $\nabla_\alpha \beta$ is an ordinary vector field that we denote it by $X_{\alpha\beta}$. Clearly, $(\alpha, \beta) \mapsto X_{\alpha\beta}$ is bilinear. A super connection on $M$ is determined by an ordinary connection on $M$, and some tensors $\omega_0, \omega_\alpha$, and $X_{\alpha\beta}$ as described above.

The (super)torsion of a super connection on $M$, is a $\hat{T}M$-valued (super) differential form of order two, and is defined as follows.

$$T(U, V) = \nabla_U V - (-1)^{UV} \nabla_V U - [U, V] \quad (15)$$

Since, $T$ preserve parities of vector fields ($|T(U,V)| = |U| + |V|$), it is an even tensor. The super torsion of a super connection is zero, if and only if its associated ordinary connection is torsion free and

$$\omega_0(U, \alpha) = \omega_\alpha(U), \quad X_{\alpha\beta} = -X_{\alpha\beta} \quad (16)$$

A torsion free super connection $\nabla$ on $M$ is determined by a triplet $\nabla = (\dot{\nabla}, \omega_0, X_{\alpha\beta})$ in which $\dot{\nabla}$ a torsion free ordinary connection on $M$, and $\omega_0$ is a tensor in $A^1(M, L(W))$, and $X_{\alpha\beta}$ are vector fields such that with respect to $\alpha, \beta$ is $C^\infty(M)$-bilinear and alternating. In this case, the covariant derivations between odd and even vector fields are as follow.

$$\nabla_U \alpha = U(\alpha) + \omega_0(U, \alpha), \quad \nabla_\alpha U = \omega_0(U, \alpha), \quad \nabla_\alpha \beta = X_{\alpha\beta} \quad (17)$$

**Super curvature:** Super curvature of a super connection, naturally is defined as follows.

$$R(U,V)(W) = \nabla_U \nabla_V W - (-1)^{UV} \nabla_V \nabla_U W - \nabla_{[U,V]} W \tag{18}$$

This is a super tensor that preserves parity and is super alternating with respect to $U$ and $V$. In other words $R$ is a $L(\hat{T}M)$-valued super differential form of order two.

$$R(U,V)(W) = -(-1)^{UV} R(V,U)(W) \tag{19}$$

**Super metrics:** An even numerical valued covariant super tensor of order two on $M$, such as $<.,.>: \hat{\mathfrak{X}}M \times \hat{\mathfrak{X}}M \to C^\infty(M)$ is called a super metric on $M$, if for each $p \in M$ it is a super inner product on $\hat{T}_p M$. So, even and odd vector fields are orthogonal to each other, and for homogeneous vector fields $U$ and $V$, we have:

$$<U,V> = (-1)^{UV} <V,U> \tag{20}$$

A super connection $\nabla$ on $M$ is called compatible with a super metric $<.,.>$, whenever for each $U,V,W \in \hat{\mathfrak{X}}M$ the following relation holds.

$$U<V,W> = <\nabla_U V, W> + (-1)^{UV} <V, \nabla_U W> \tag{21}$$

The restriction of a super metric on $M$ to $TM$, is a semi-Riemannian metric on $M$. So, a super metric on $M$ is a pair $(g^e, g^o)$ in which $g^e$ is a semi-Riemannian metric on $M$ and $g^o$ is a nondegenerate symplectic form on the trivial bundle $M \times W$.

There exists a unique torsion free super connection on $M$ that is compatible with a given super metric. This connection is the super version of the Levi-civita connection and can be found by the extended Koszul formula as follows[3].

$$\begin{aligned} 2<\nabla_U V, W> = &\; U<V,W> + (-1)^{UV} V<U,W> - (-1)^{W(U+V)} W<U,V> \\ &+ <[U,V],W> - <U,[V,W]> - (-1)^{VW} <[U,W],V> \end{aligned} \tag{22}$$

If $(g^e, g^o)$ is a super metric on $M$, and $\nabla = (\dot{\nabla}, \omega_0, X_{\alpha\beta})$ is its super Levi-civita connection, then $\dot{\nabla}$ is the Levi-civita connection of $g^e$ and

$$<\omega_0(U,\alpha),\beta> = <U, X_{\beta\alpha}> = \frac{1}{2} U(g^o)(\alpha,\beta) \tag{23}$$

The curvature tensor of the super Levi-civita connection has the following properties[3].

$$<R(U,V)(Z),W> = -(-1)^{ZW} <R(U,V)(W),Z>$$
$$<R(U,V)(Z),W> = (-1)^{(U+V)(Z+W)} <R(Z,W)(U),V> \qquad (24)$$

To define super Ricci curvature and scalar curvature we need a homogeneous local base $\{E_1,\cdots,E_n,\alpha_1,\cdots,\alpha_m\}$ of $\hat{T}M$ and its reciprocal base $\{E^1,\cdots,E^n,\alpha^1,\cdots,\alpha^m\}$.

$$Ric(U,V) = \sum_i <R(U,E_i)(E^i),V> + \sum_j <R(U,\alpha_j)(\alpha^j),V> \qquad (25)$$

$$R = \sum_i Ric(E_i,E^i) + \sum_j Ric(\alpha_j,\alpha^j) \qquad (26)$$

$Ric$ is super symmetric, in other words $Ric(U,V) = (-1)^{UV} Ric(V,U)$

## 5. Curvatures

In this section, we investigate the case $\dim(W) = 2$, and we will find super levi-civita connection and its curvature explicitly with respect to the symmetric and symplectic part of the super metric. We fix a nondegenrate symplectic product $\eta$ on $W$. Any symplectic metric on $M \times W$ is $\pm e^{2\theta}\eta$ for some function $\theta \in C^\infty(M)$. By changing $\eta$ to $-\eta$, if necessary, we can assume any super metric on $M$ has the form $(g,e^{2\theta}\eta)$ in which $g$ is a semi Riemannian metric on $M$ and $\theta \in C^\infty(M)$. So, super metrics on $M$ are equivalent to the pairs $(g,\theta)$.

**Theorem:** If $\boldsymbol{g} = (g,\theta)$ is a super metric on $M$ and $\nabla = (\dot{\nabla},\omega_0,X_{\alpha\beta})$ is its super Levi-civita connection, then

$$\omega_0(U,\alpha) = U(\theta)\alpha, \; X_{\alpha\beta} = <\beta,\alpha>\vec{\nabla}\theta \qquad (27)$$

**Proof:** Assume $\alpha,\beta \in C^\infty(M,W)$ are constant odd vector fields,

$$<\omega_0(U,\alpha),\beta> = \frac{1}{2}U(g^o)(\alpha,\beta) = \frac{1}{2}U\left(e^{2\theta}\eta(\alpha,\beta)\right)$$
$$= U(\theta)e^{2\theta}\eta(\alpha,\beta) = <U(\theta)\alpha,\beta> \implies \omega_0(U,\alpha) = U(\theta)\alpha \qquad (28)$$

Clearly, we can deduce that this equality also holds for arbitrary odd fields.

Now, we find the other equality.

$$<U, X_{\beta\alpha}> = <\omega_0(U,\alpha), \beta> = <U(\theta)\alpha, \beta> = U(\theta)<\alpha, \beta>$$
$$= <U, \vec{\nabla}\theta><\alpha, \beta> = <U, <\alpha, \beta>\vec{\nabla}\theta> \implies X_{\beta\alpha} = <\alpha, \beta>\vec{\nabla}\theta \quad \bullet \tag{29}$$

This theorem implies that

$$\nabla_\alpha U = U(\theta)\alpha \tag{30}$$
$$\nabla_U \alpha = U(\alpha) + U(\theta)\alpha \tag{31}$$
$$\nabla_\alpha \beta = <\beta, \alpha>\vec{\nabla}\theta \tag{32}$$

We remind that Hessian and Laplacian of $\theta$ is defined as follows [4].

$$Hes(\theta)(U,V) = (\nabla_U d\theta)(V) = U(V(\theta)) - (\nabla_U V)(\theta) \tag{33}$$

$$\Delta(\theta) = \text{div}(d\theta) = \sum_{i=1}^n Hes(\theta)(E_i, E^i) \tag{34}$$

It is convenient to define the symmetric tensor $S_\theta$ as follows.

$$S_\theta = Hes(\theta) + d\theta \otimes d\theta \tag{35}$$

Contraction of $S_\theta$ is $\Delta(\theta) + |\vec{\nabla}\theta|^2$.

**Theorem:** The super curvature of the super Levi-civita connection of a super metric $(g, \theta)$ on $M$ satisfies the following equalities.

$$\mathbf{R}(U,V)(W) = R(U,V)(W) \tag{36}$$
$$\mathbf{R}(U,V)(\alpha) = 0 \tag{37}$$
$$\mathbf{R}(U,\alpha)(V) = S_\theta(U,V)\alpha \tag{38}$$
$$\mathbf{R}(U,\alpha)(\beta) = <\beta, \alpha>\left(U(\theta)\vec{\nabla}\theta + \nabla_U(\vec{\nabla}\theta)\right) \tag{39}$$
$$\mathbf{R}(\alpha,\beta)(U) = 0 \tag{40}$$
$$\mathbf{R}(\alpha,\beta)(\gamma) = |\vec{\nabla}\theta|^2(<\gamma,\alpha>\beta + <\gamma,\beta>\alpha) \tag{41}$$

**Proof:** All these equalities are proved by straight computations. For example, we prove (38). Assume $\alpha$ is constant.

$$\begin{aligned}\mathbf{R}(U,\alpha)(V) &= \nabla_U \nabla_\alpha V - \nabla_\alpha \nabla_U V = \nabla_U(V(\theta)\alpha) - (\nabla_U V)(\theta)\alpha \\ &= U(V(\theta))\alpha + V(\theta)U(\theta)\alpha - (\nabla_U V)(\theta)\alpha \\ &= Hes(\theta)(U,V)\alpha + (d\theta \otimes d\theta)(U,V)\alpha = S_\theta(U,V)\alpha \quad \bullet\end{aligned} \tag{42}$$

**Theorem:** The super Ricci curvature of the super Levi-civita connection of a super metric $(g, \theta)$ on $M$ satisfies the following equalities.

$$\mathbf{Ric}(U,V) = \text{Ric}(U,V) + 2S_\theta(U,V) \tag{43}$$
$$\mathbf{Ric}(U,\alpha) = 0 \tag{44}$$
$$\mathbf{Ric}(\alpha,\beta) = <\beta,\alpha> \left(\Delta(\theta) - 2|\vec{\nabla}\theta|^2\right) \tag{45}$$

**Proof:** All these equalities are also proved by straight computations. Consider a homogeneous local base $\{E_1, \cdots, E_n, \alpha_1, \alpha_2\}$ of $\hat{T}M$ and its reciprocal base $\{E^1, \cdots, E^n, \alpha^1, \alpha^2\}$. Proof of (43):

$$\begin{aligned}
\mathbf{Ric}(U,V) &= \sum_i <R(U,E_i)(E^i),V> + \sum_j <R(U,\alpha_j)(\alpha^j),V> \\
&= \sum_i <R(U,E_i)(E^i),V> + \sum_j <\alpha^j,\alpha_j><U(\theta)\vec{\nabla}\theta + \nabla_U(\vec{\nabla}\theta),V> \\
&= \text{Ric}(U,V) + 2S_\theta(U,V)
\end{aligned} \tag{46}$$

Proof of (45):

$$\begin{aligned}
\mathbf{Ric}(\alpha,\beta) &= \sum_i <R(\alpha,E_i)(E^i),\beta> + \sum_j <R(\alpha,\alpha_j)(\alpha^j),\beta> \\
&= \sum_i -<S_\theta(E_i,E^i)\alpha,\beta> + \sum_j |\vec{\nabla}\theta|^2 <<\alpha^j,\alpha>\alpha_j + <\alpha^j,\alpha_j>\alpha,\beta> \\
&= -\left(\Delta(\theta) + |\vec{\nabla}\theta|^2\right)<\alpha,\beta> + |\vec{\nabla}\theta|^2 \sum_j \left(<\alpha^j,\alpha><\alpha_j,\beta> + <\alpha^j,\alpha_j><\alpha,\beta>\right) \\
&= -\left(\Delta(\theta) + |\vec{\nabla}\theta|^2\right)<\alpha,\beta> + |\vec{\nabla}\theta|^2 \left(-<\beta,\alpha> + 2<\alpha,\beta>\right) \\
&= <\beta,\alpha>\left(\Delta(\theta) - 2|\vec{\nabla}\theta|^2\right)
\end{aligned} \tag{47}$$

●

**Theorem:** The scalar curvature of the super Levi-civita connection of a super metric $(g, \theta)$ on $M$ is as follows.

$$\mathbf{R} = R + 4\Delta(\theta) - 2|\vec{\nabla}\theta|^2 \tag{48}$$

**Proof:**

$$\begin{aligned}
\mathbf{R} &= \sum_i \mathbf{Ric}(E_i,E^i) + \sum_j \mathbf{Ric}(\alpha_j,\alpha^j) \\
&= \sum_i \left(\text{Ric}(E_i,E^i) + 2S_\theta(E_i,E^i)\right) + \sum_j <\alpha^j,\alpha_j>\left(\Delta(\theta) - 2|\vec{\nabla}\theta|^2\right) \\
&= R + 2\left(\Delta(\theta) + |\vec{\nabla}\theta|^2\right) + 2\left(\Delta(\theta) - 2|\vec{\nabla}\theta|^2\right) \\
&= R + 4\Delta(\theta) - 2|\vec{\nabla}\theta|^2
\end{aligned} \tag{49}$$

●

## 6. Super spacetime and field equation

In this section, we use the same method as in [4]. Consider $M$ as a space-time manifold whose dimension is $n$ and $2 \leq n$. Let $\boldsymbol{g} = (g, \theta)$ is a super metric on $M$. So, $g$ is a semi-Riemannian metric on $M$ and play the role of the potential for gravity. We will find that $\theta$ plays the role of potential for some mass distribution in space-time that can be interpreted as dark matter. In this structure, even part of $\hat{T}M$ models gravity and its odd part models some mass distribution on space-time that is a good candidate for dark matter.

To find a proper field equation, we use Hilbert action in the context of super metrics on $M$. Denote the canonical volume form of a metric $g$ on oriented manifold $M$ by $\Omega_g$. Hilbert action $L$ is defined as follows.

$$L(g, \theta) = \int_M \boldsymbol{R} \Omega_g = \int_M \left( R + 4\Delta(\theta) - 2|\vec{\nabla}\theta|^2 \right) \Omega_g \qquad (50)$$

To be more precise, we must integrate on open subsets $U$ of $M$ such that $\bar{U}$ is compact [2].

A variation for a super metric $\boldsymbol{g}$ is obtained by a pair $(s, h)$ in which $s$ is a symmetric covariant tensor of order two on $M$ and $h \in C^\infty(M)$. Set

$$g(t) = g + ts, \quad \theta(t) = \theta + th \qquad (51)$$

For small $t$, $\boldsymbol{g}(t) = (g(t), \theta(t))$ is a super metric on $M$ and is a variation of $\boldsymbol{g}$. A super metric $\boldsymbol{g}$ is a critical metric for the Hilbert action iff for every pair $(s, h)$:

$$\left.\frac{d}{dt}\right|_{t=0} L(g(t), \theta(t)) = \int_M \left.\frac{d}{dt}\right|_{t=0} \boldsymbol{R}(t) \Omega_{g(t)} = 0 \qquad (52)$$

In above, $\boldsymbol{R}(t)$ is scalar curvature of $\boldsymbol{g}(t)$, so

$$\boldsymbol{R}(t) = R(t) + 4\Delta^t(\theta(t)) - 2|\vec{\nabla}^t \theta(t)|^2 \qquad (53)$$

To find $\left.\frac{d}{dt}\right|_{t=0} \boldsymbol{R}(t)\Omega_{g(t)}$ we must find

$$R'(0), \ \Delta^t(\theta(t))'(0), \ \left(|\vec{\nabla}^t\theta(t)|^2\right)'(0), \ (\Omega_{g(t)})'(0) \qquad (54)$$

In [4], these derivations were computed as follows.

$$R'(0) = -<s, \text{Ric}> + \text{div}(W), \quad W \in \mathfrak{X}M \tag{55}$$

$$(\Omega_{g+ts})'(0) = \frac{1}{2}<g,s>\Omega_g \tag{56}$$

$$(\Delta^t(\theta(t)))'(0) = \Delta(h) - \text{div}(s(\vec{\nabla}\theta) + \frac{1}{2}<\vec{\nabla}\text{tr}(s), \vec{\nabla}\theta> \tag{57}$$

$$\left(|\vec{\nabla}^t\theta(t)|^2\right)'(0) = -<s, d\theta\otimes d\theta> + 2<\vec{\nabla}h, \vec{\nabla}\theta> \tag{58}$$

Note that the integral of divergence of every vector fields on $M$ is zero, so integral of Laplacian of any smooth function on $M$ is zero. Also, $\text{tr}(s) = <s,g>$ and for any two smooth function $f$ and $h$ we have [2]

$$\int_M <\vec{\nabla}f, \vec{\nabla}h>\Omega_g = -\int_M f\Delta(h)\Omega_g \tag{59}$$

Now, we can compute critical super metric for Hilbert action.

$$\begin{aligned}
\frac{d}{dt}\bigg|_{t=0} L(g(t),\theta(t)) &= \int_M \frac{d}{dt}\bigg|_{t=0}(R(t) + 4\Delta^t(\theta(t)) - 2|\vec{\nabla}^t\theta(t)|^2)\Omega_{g(t)} \\
&= \int_M (R'(0) + 4\left(\Delta^t(\theta(t))\right)'(0) - 2\left(|\vec{\nabla}^t\theta(t)|^2\right)'(0))\Omega_g \\
&\quad + \int_M \left(R + 4\Delta(\theta) - 2|\vec{\nabla}\theta|^2\right)(\Omega_{g(t)})'(0) \\
&= \int_M (-<s, \text{Ric}> + \text{div}(W) + 4\left(\Delta(h) - \text{div}(s(\vec{\nabla}\theta) + \frac{1}{2}<\vec{\nabla}\text{tr}(s), \vec{\nabla}\theta>\right) \\
&\quad - 2(-<s, d\theta\otimes d\theta> + 2<\vec{\nabla}h, \vec{\nabla}\theta>))\Omega_g \\
&\quad + \int_M \left(R + 4\Delta(\theta) - 2|\vec{\nabla}\theta|^2\right)\frac{1}{2}<g,s>\Omega_g \\
&= \int_M (<s, -\text{Ric}> - 2\text{tr}(s)\Delta(\theta) + <s, 2d\theta\otimes d\theta> \\
&\quad + <s, \frac{1}{2}\left(R + 4\Delta(\theta) - 2|\vec{\nabla}\theta|^2\right)g>\Omega_g + 4\int_M h\Delta(\theta)\Omega_g \\
&= \int_M (<s, -\text{Ric} + 2d\theta\otimes d\theta + \frac{1}{2}\left(R - 2|\vec{\nabla}\theta|^2\right)g>\Omega_g + 4\int_M h\Delta(\theta)\Omega_g
\end{aligned} \tag{60}$$

The above expression is zero for all pair $(s,h)$ iff

$$\text{Ric} - \frac{1}{2}Rg = 2d\theta\otimes d\theta - |\vec{\nabla}\theta|^2 g \tag{61}$$

$$\Delta(\theta) = 0 \tag{62}$$

These are exactly the same equations we have found, by some other structure, in [4]. Assume $\vec{\nabla}\theta$ is timelike, so $\vec{\nabla}\theta$ can be interpreted as the current of some mass and (62) shows the conservation of this mass. Here, $\rho = \sqrt{-|\vec{\nabla}\theta|^2}$ is the density of the mass. Right hand side of (61) is the momentum-energy tensor of this mass, so momentum-energy tensor of this mass is also conserved.

Note that equation (61) implies (62). Because, divergence of the left hand side of (61) is zero, so divergence of the right hand side of (61) is zero. But divergence of $2d\theta \otimes d\theta - |\vec{\nabla}\theta|^2 g$ is $2(\Delta\theta)d\theta$ [4]. So, $(\Delta\theta)d\theta = 0$ and this equation implies $\Delta\theta = 0$. Therefore, equation (61) is sufficient for determining relation between $g$ and .

In the case $3 \leq n$, by computing contractions of both side of (61) we find that $R = 2|\vec{\nabla}\theta|^2$. So (61) is equivalent to:

$$\text{Ric} = 2d\theta \otimes d\theta \qquad (63)$$

If at every point of $M$, $\vec{\nabla}\theta \neq 0$ and $\vec{\nabla}\theta$ is timelike, then the level sets of $\theta$ (defined by $L_c = \theta^{-1}(c)$) are $n-1$ dimensional spacelike submanifolds of $M$ that can be considered as a natural simultaneity for events. Integral curves of $\vec{\nabla}\theta$ pass through level sets of $\theta$, orthogonally. For two distinct level set $L_c$ and $L_d$, suppose every integral curve of $\vec{\nabla}\theta$ that intersect $L_c$ also intersect $L_d$. So, we find a one to one correspondence between $L_c$ and $L_d$. Under this correspondence, any open set of $L_c$ corresponds to an open set of $L_d$.

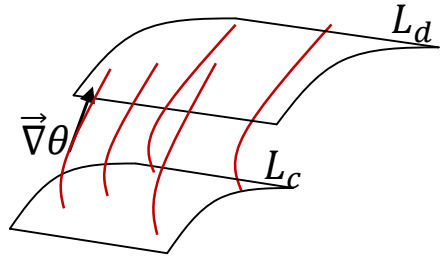

The submanifolds $L_c$ and $L_d$ are Riemannian manifolds and have canonical volume forms $\Omega_c$ and $\Omega_d$ respectively. For an open set $W \subseteq L_c$ whose closure is compact, its corresponding open set $W'$, has compact closure too. Relation $\Delta(\theta) = 0$ implies the following equality holds that means the conservation of mass.

$$\int_W \rho\Omega_c = \int_{W'} \rho\Omega_d \tag{64}$$

**7.Some Solution**

We can solve field equation to find some space-time satisfying field equation. Suppose that $(N, g)$ is a Riemannian manifold with dimension $n$ and $2 \leq n$ and $M = N \times (0, +\infty)$. Tangent vectors to $M$ at $(p, t)$ is of the form $(v, (t, \lambda))$ in which $v \in T_pN$ and $\lambda \in \mathbb{R}$. Denote the vector field $(0, (t, 1))$ on $M$ by $\partial_t$. For a smooth function $f(p, t)$, we have $\partial_t(f) = \frac{\partial f}{\partial t}$.

We denote vector fields on $N$ by $X, Y, Z, \ldots$ and consider them as special vector fields on $M$ that do not depend on . Denote the map $(p, t) \mapsto t$ on $M$ by $t$. For the 1-form $dt$ we have $dt(v, (s, \lambda)) = \lambda$, so $dt(\partial t) = 1$ and $dt(X) = 0$. For some smooth functions $h, k : (0, +\infty) \to \mathbb{R}$ and $f : N \to \mathbb{R}$ define a metric $\hat{g} = <,>$ on $M$ as follows.

$$\hat{g} = e^{2h}g - e^{2k+2f} dt \otimes dt \tag{65}$$

By this definition, we have:

$$<X, Y> = e^{2h}g(X, Y), \quad <X, \partial_t> = 0, \quad <\partial_t, \partial_t> = -e^{2k+2f} \tag{66}$$

Consider $\theta : N \times (0, +\infty) \to \mathbb{R}$ such that it depends only on $t$ so $\theta(p, t) = \theta(t)$. In other words, level sets of $\theta$ are $N \times \{c\}$. We find $h, k, f, \theta$ such that $(\hat{g}, \theta)$ is a solution for the field equations (63).

These assumptions imply that :

$$d\theta = \theta' dt, \quad \vec{\nabla}\theta = -e^{-2k-2f}\theta'\partial_t, \quad |\vec{\nabla}\theta|^2 = -e^{-2k-2f}\theta'^2 \tag{67}$$

Denote the Levi-Civita connections of $M$ and $N$ by $\nabla$ and $\nabla^N$ respectively. In the following formulas, gradient and Laplacian of $f$ and norms are with respect to Riemannian structure of $N$. Straightforward computations show that:

$$\nabla_X Y = \nabla_X^N Y + e^{-2k-2f} h' <X,Y> \partial_t \qquad (68)$$
$$\nabla_{\partial_t} Y = h'Y + X(f)\partial_t \qquad (69)$$
$$\nabla_{\partial_t} \partial_t = e^{2k+2f} \overrightarrow{\nabla^N} f + k'\partial_t \qquad (70)$$

First, we find $\Delta(\theta)$. Suppose $\{E_1, \cdots, E_n\}$ is a local base on $N$ and $\{E^1, \cdots, E^n\}$ is its reciprocal base. So, $\{e^{-2h}E^1, \cdots, e^{-2h}E^n, -e^{-2k-2f}\partial_t\}$ is the reciprocal base of the $\{E_1, \cdots, E_n, \partial_t\}$ that is a local base of $M$. Laplacian of $\theta$ can be computed as follows.

$$\begin{aligned}
\Delta(\theta) &= \sum_i <\nabla_{E_i}\vec{\nabla}\theta, e^{-2h}E^i> + <\nabla_{\partial_t}\vec{\nabla}\theta, -e^{-2k-2f}\partial_t> \\
&= \sum_i e^{-2h} <\nabla_{E_i}(-e^{-2k-2f}\theta'\partial_t), E^i> - e^{-2k-2f} <\nabla_{\partial_t}(-e^{-2k-2f}\theta'\partial_t), \partial_t> \\
&= \sum_i \left(-e^{-2h}e^{-2k-2f}\theta' <\nabla_{E_i}\partial_t, E^i>\right) \\
&\quad + e^{-2k-2f}e^{-2k-2f}(\theta'' - 2\theta'k') <\partial_t, \partial_t> + e^{-2k-2f}e^{-2k-2f}\theta' <\nabla_{\partial_t}\partial_t, \partial_t> \\
&= \sum_i \left(-e^{-2h}e^{-2k-2f}\theta' <h'E_i, E^i>\right) - e^{-2k-2f}(\theta'' - 2\theta'k') - e^{-2k-2f}\theta'k' \\
&= -e^{-2k-2f}(nh'\theta' - k'\theta' + \theta'')
\end{aligned} \qquad (71)$$

Set $H = nh - k$. Equation $\Delta(\theta) = 0$ is equivalent to $H'\theta' + \theta'' = 0$. This equation implies that $\theta' = ce^{-H} = ce^{-nh+k}$ (for some $c$).

Now, we compute the curvature tensors of $\hat{g}$. Denote the curvature tensors and Ricci curvature tensors of $M$ and $N$ by $R, R^N, \text{Ric}, \text{Ric}^N$ respectively. Straight forward computations show that:

$$\begin{aligned}
<R(X,Y)(Z), W> &= <R^N(X,Y)(Z), W> \\
&\quad + e^{-2k-2f}h'^2(<Y,Z><X,W> - <X,Z><Y,W>)
\end{aligned} \qquad (72)$$

$$\begin{aligned}
<R(\partial_t, X)(Y), \partial_t> &= <X,Y>(h'k' - h'^2 - h'') \\
&\quad + e^{2k+2f}(X(f)Y(f) + \text{Hes}(f)(X,Y))
\end{aligned} \qquad (73)$$

$$\begin{aligned}
\text{Ric}(X,Y) &= \text{Ric}^N(X,Y) - X(f)Y(f) - \text{Hes}(f)(X,Y) \\
&\quad + e^{-2k-2f} <X,Y>(h'' + nh'^2 - h'k')
\end{aligned} \qquad (74)$$

$$\text{Ric}(\partial_t, \partial_t) = e^{-2h+2k+2f}\left(\left|\overrightarrow{\nabla^N}f\right|^2 + \Delta^N(f)\right) + n(h'k' - h'' - h'^2) \qquad (75)$$

Equation $\text{Ric} = 2d\theta \otimes d\theta = 2\theta'^2 dt \otimes dt$ holds if and only if:

$$0 = Ric^N(X,Y) - X(f)Y(f) - Hes(f)(X,Y)$$
$$+ e^{-2k-2f} <X,Y> (h'' + nh'^2 - h'k') \tag{76}$$
$$2\theta'^2 = e^{-2h+2k+2f}\left(\left|\overrightarrow{\nabla^N}f\right|^2 + \Delta^N(f)\right) + n(h'k' - h'' - h'^2) \tag{77}$$

Equation (76) implies

$$e^{2f}\frac{Ric^N(X,Y) - X(f)Y(f) - Hes(f)(X,Y)}{g(X,Y)} = -e^{-2k+2h}(h'' + nh'^2 - h'k') \tag{78}$$

Left hand side of (78) does not depend on $t$ and right hand side of (78) only depends on $t$, so for some scalar $\lambda$ we have:

$$Ric^N(X,Y) - X(f)Y(f) - Hes(f)(X,Y) = \lambda e^{-2f} g(X,Y) \tag{79}$$
$$-e^{-2k+2h}(h'' + nh'^2 - h'k') = \lambda \tag{80}$$

For $\lambda \neq 0$ there exists no solution but for $\lambda = 0$, we have some solutions. Assume $\lambda = 0$, so, $(N,g)$ must be a Riemannian manifold that its Ricci curvature satisfies the following relation for some smooth function $f$.

$$Ric^N(X,Y) = X(f)Y(f) + Hes(f)(X,Y) \tag{81}$$

But, equation (77) implies

$$e^{2h-2k}\left(2\theta'^2 - n(h'k' - h'' - h'^2)\right) = e^{2f}\left(\left|\overrightarrow{\nabla^N}f\right|^2 + \Delta^N(f)\right) \tag{82}$$

Right hand side of (82) does not depend on $t$ and left hand side of (82) only depends on $t$, so for some scalar $\lambda$ we have:

$$e^{2h-2k}\left(2\theta'^2 - n(h'k' - h'' - h'^2)\right) = \lambda \tag{83}$$
$$e^{2f}\left(\left|\overrightarrow{\nabla^N}f\right|^2 + \Delta^N(f)\right) = \lambda \tag{84}$$

For $\lambda = 0$, we find $\left|\overrightarrow{\nabla^N}f\right|^2 + \Delta^N(f) = 0$, but (80) implies that $\left|\overrightarrow{\nabla^N}f\right|^2 + \Delta^N(f)$ is the scalar curvature of $(N,g)$. So, $(N,g)$ must also have zero scalar curvature. Now, we must solve the following equations.

$$h'' + nh'^2 - h'k' = 0 \tag{85}$$
$$h'k' - h'' - h'^2 = \frac{2c^2}{n} e^{-2nh+2k} \tag{86}$$

Set $K(t) = \int_0^t e^{k(x)} dx$. Solving these equation yields:

$$h = \frac{1}{n}\ln(K), \quad \theta = \sqrt{\frac{n-1}{2n}} \ln K, \quad k = \ln K' \tag{87}$$

So, for any smooth positive increasing function $K: [0, +\infty) \to \mathbb{R}$ such that $K(0) = 0$, we have a solution for field equations (63). At any point $\in N$, consider the observer $\alpha(t) = (p, t)$ that rest at $p$. Let $\tau(t, p)$ be the proper time measured by this observer, from the beginning of the universe (Big Bang) until time the universe reached at $N \times \{t\}$. Since, $\alpha'(t) = (\partial_t)_{(p,t)}$, we have:

$$\tau(t,p) = \int_0^t \sqrt{-\langle \partial_{t(p,t)}, \partial_{t(p,t)} \rangle}\, dt = \int_0^t \sqrt{e^{2k(t)+2f(p)}}\, dt$$
$$= \int_0^t e^{k(t)+f(p)} dt = e^{f(p)} \int_0^t K' dt = e^{f(p)} K(t) \tag{88}$$

The density of dark matter at $(p, t)$ is

$$\rho = \sqrt{-|\vec{\nabla}\theta|^2} = e^{-k-f}\theta' = e^{-k-f}\sqrt{\frac{n-1}{2n}} \frac{e^k}{K} = \sqrt{\frac{n-1}{2n}} \frac{1}{e^f K} = \sqrt{\frac{n-1}{2n}} \frac{1}{\tau(t,p)} \tag{89}$$

So,

$$(t, p) = \sqrt{\frac{n-1}{2n}} \frac{1}{\rho}$$

This relation means the proper time measured by the observer $\alpha(t)$ is proportional to the inverse of density of dark matter. Heavier mass cause time goes slower.

[3]  Boroojerdian N., (2020) *Geometric Structures and Differential Operators on Manifolds Having Super tangent Bundle*, arXiv:2011.07382

[4] Boroojerdian N., (2013) *Geometrization of mass in general relativity*. Int J Theor Phys 52(7):2432

[5]  Claudio Carmeli, Lauren Caston, Rita Fioresi, *Mathematical Foundations of Supersymmetry*, European Mathematical Society, 2011

[6]  S. James Gates, Marcus T. Grisaru, Martin Roˇcek, Warren Siegel, *SUPERSPACE.or One thousand and one lessons in supersymmetry*, ISBN 0-8053-3160-3